\documentclass[12pt,preprint]{emulateapj}
\usepackage{graphicx}

\shortauthors{Dragomir et al.}
\begin{document}

\title{A search for transits of GJ 581\lowercase{e} and characterization of the host star variability using MOST\footnotemark[*] space telescope photometry}\footnotetext[*]{Based on data from the MOST satellite, a Canadian Space Agency mission operated by Microsatellite Systems Canada Inc. (MSCI; former Dynacon Inc.) and the Universities of Toronto and British Columbia, with the assistance of the University of Vienna.}

\author{
  Diana Dragomir\altaffilmark{1},
  Jaymie M. Matthews\altaffilmark{1},
  Rainer Kuschnig\altaffilmark{2},
  Jason F. Rowe\altaffilmark{3},
  Brett J. Gladman\altaffilmark{1},
  David B. Guenther\altaffilmark{4},
  Anthony F. J. Moffat\altaffilmark{5},
  Slavek M. Rucinski\altaffilmark{6},
  Dimitar Sasselov\altaffilmark{7}, 
  Werner W. Weiss\altaffilmark{2}
}

\email{diana@phas.ubc.ca}
\altaffiltext{1}{Department of Physics and Astronomy, University of
  British Columbia, Vancouver, BC V6T1Z1, Canada}
\altaffiltext{2}{Universit\"{a}t Wien, Institut f\"{u}r Astronomie, 
 T\"{u}rkenschanzstrasse 17, AÐ1180 Wien, Austria}
\altaffiltext{3}{NASA Ames Research Center, Moffett Field, CA 94035}
\altaffiltext{4}{Department of Astronomy and Physics, St. MaryÕs University Halifax, NS B3H 3C3, Canada}
\altaffiltext{5}{D\'{e}pt	 de physique, Univ de Montr\'{e}al C.P. 6128, Succ. Centre-Ville, Montr\'{e}al, QC H3C 3J7, and Obs du mont M\'{e}gantic, Canada} 
\altaffiltext{6}{Department of Astronomy and Astrophysics, University of Toronto, 50 St. George Street, Toronto, ON M5S 3H4, Canada}
\altaffiltext{7}{Harvard-Smithsonian Center for Astrophysics, 60 Garden Street, Cambridge, MA 02138, USA}

%%%%%%%%%%%%%%%%%%%%%%%%%%%%%%%%%%%%%%%%%%%%%%%%%%%%%%%%%%%%%%%%%%%%

\begin{abstract}

The GJ 581 system has been amply studied since its discovery in 2005: the number of known planets in the system has increased and their orbital parameters are among the most precisely determined for radial velocity detected exoplanets. We have acquired MOST space-based photometry during 2007 and 2009, with the aims of measuring the stellar variability and searching for transits of GJ 581e, respectively. We quantify our sensitivity to shallow transit signals using Monte Carlo simulations, and perform a transit search within the 3$\sigma$ transit windows corresponding to both the circular and Keplerian orbit ephemerides. Our analysis rules out transits for a planet with an orbital period of 3.15 days (GJ 581 e) having a radius larger than 1.62 $R_{\oplus}$ (or a density lower than 2.39 g cm$^{-3}$ for an orbital inclination of 90$^{\circ}$) to 2$\sigma$ confidence. Thus, if the planet transits, we can exclude hydrogen, helium and water theoretical model compositions. The MOST photometry also allows us to rule out transits of GJ 581b within the Keplerian orbit-derived transit window for impact parameter values smaller than $\sim$0.4 and confirm previous results which exclude transits for this planet within the circular orbit-derived transit window, for all plausible interior compositions. We find that the stellar brightness of GJ 581 is stable to within 1$\%$, a characteristic which is favourable to the development of life in the habitable zone of the system. In the 2009 photometry, we detect a stellar signal with a period of 5.586 $\pm$ 0.051 days, which is close to the orbital period of GJ 581b ($P=$5.37 days). However, further monitoring of the system is necessary to verify the nature of this variation.

\end{abstract}

\keywords{planetary systems -- techniques: photometric -- stars: individual (GJ~581)}

\section{Introduction}

Radial velocity (RV) and transit searches are discovering progressively less massive and smaller exoplanets. The HARPS \citep{May03}, Keck \citep{But96} several other RV surveys have brought forth dozens of planets with minimum masses $<$10 M$_{\oplus}$, while the {\it Kepler} mission \citep{Bor10} has so far yielded a wealth of super-Earth candidates with radii $<$4 R$_{\oplus}$. Furthermore, Kepler has already detected three planetary candidates with radii smaller than that of the Earth in a single system \citep{Mui12}. 

There are currently a handful of super-Earths for which both the masses and radii have been measured. By determining the density of these planets, of which none exist in the Solar system, it becomes possible to constrain their composition and eventually their formation. Thanks to their transiting nature it also becomes possible to perform spectroscopic studies of these planets. Such studies, rendered possible by the large transit depth which is induced by the small size of the host star relative to that of the transiting planet, now exist for the super-Earth GJ 1214b (\citealt{Bea10}, \citealt{Bea11}, \citealt{Cro11}, \citealt{Ber12}). 

GJ 581, a small M2.5 dwarf ($R_{\star}$=0.299 R$_{\odot}$; \cite{Bra11}), is known to host up to 6 planets (\citealt{Bon05}, \citealt{Udr07}, \citealt{May09}, \citealt{Vog10}). However, we note that the existence of planets f and g has been disputed and arguably ruled out by several independent analyses (\citealt{Gre11}, \citealt{For11}, \citealt{Tuo11}). This system has been observed and studied at length. The published HARPS radial velocity measurements span approximately seven years \citep{For11}, and the HIRES observations cover more than a decade. The orbital parameters of four planets in the system are well constrained \citep{For11}. At least three of those planets (c, d and e) have minimum masses in the super-Earth regime. The habitability of c and d has been assessed by several groups, and while the former orbits too close to the star to sustain liquid water on its surface (\citealt{Blo07}, \citealt{Sel07}), GJ 581 d was found to reside in the habitable zone of the system (\citealt{Blo07}, \citealt{Wor11}, \citealt{Bar09}, \citealt{Par11}, \citealt{Kal11}).

The GJ 581 system was observed by the {\it MOST} ({\it Microvariability and Oscillations of STars}; \citealt{Wal03}, \citealt{Mat04}) space telescope in order to determine its variability and search for transits of GJ 581e. These observations were acquired as part of a program aiming to search for transits of known RV super-Earth candidates using MOST. The telescope is optimized for bright star (V $<$ 10) photometry, a characteristic it shares with radial velocity surveys. Even though GJ 581 is slightly fainter (V = 10.6) than the usual MOST targets, the opportunity to obtain continuous coverage of more than one orbital cycle of the planet (thus eliminating the chance that a transit might be missed) motivated the observations. 

GJ 581e, the innermost planet in the system, orbits its host star with a period of 3.15 days and has a {\it a priori} transit probability of 5$\%$. This value corresponds to the geometric probability that the planet crosses the disk of the star and in the case of a circular orbit it is derived from the planet's semi-major axis and the stellar radius. GJ 581e, whose published properties can be found in Table 1, is the planet with the smallest radial velocity-derived minimum mass to date. If the planet is found to transit, its size can be determined and spectroscopic studies of its atmosphere become possible. 

The duration and nearly contiguous nature of the MOST observations presented in this paper also invite a search for transits of GJ 581b ($P = 5.37$ days), regardless of the outcome of our GJ 581e transit search. Indeed, even if GJ 581e does not transit, GJ 581b may still transit if their orbits are not co-planar. Before the existence of the other three planets in the system was known, \cite{Lop06} had ruled out transits of GJ 581b using an orbital solution based on the assumption of a circular orbit. However, when the three subsequently discovered planets are taken into account and their orbital eccentricities are allowed to vary during the fitting process, the resulting solution shifts the predicted mid-transit time of GJ 581b outside the observations acquired by \cite{Lop06}, as was determined by \cite{For11}. The published orbital parameters of GJ 581b are found in Table 2.

\begin{deluxetable}{ccc}
  \tablecaption{\label{planet} Published Orbital Parameters for GJ 581\lowercase{e} from \cite{For11}}
  \tablewidth{7cm}
  \tablehead{
    \colhead{Parameter} &
    \colhead{Circular model} & 
    \colhead{Keplerian model} 
  }
  \startdata
  $P$ (days)                       & $3.14941 \pm  0.00022 $ & $3.14945 \pm 0.00017$   \\
  $T_0\,^{a}$ (HJD)  & $2454749.026 \pm 0.056$  &  $2454750.31 \pm 0.13$ \\
  $e$                              & $-$  &  $0.32 \pm 0.09$  \\
  $\omega$ (deg)                   & $-$ & $ 236 \pm 17 $ \\
  $K$ (m\,s$^{-1}$)             & $1.754 \pm 0.180$  & $1.96 \pm 0.20 $ \\
  $M_p \sin i$ ($M_{\oplus}$)             & $1.84 $ & $ 1.95 $ \\
  $a$ (AU)                         & $0.028$  &  $ 0.028 $ \\
  \enddata
  \tablenotetext{a}{Predicted mid-transit time.}
\end{deluxetable}

\begin{deluxetable}{ccc}
  \tablecaption{\label{planet} Published Orbital Parameters for GJ 581\lowercase{b} from \cite{For11}}
  \tablewidth{7cm}
  \tablehead{
    \colhead{Parameter} &
    \colhead{Circular model} & 
    \colhead{Keplerian model} 
  }
  \startdata
  $P$ (days)                       & $5.36864 \pm  0.00009 $ & $5.36865 \pm 0.00009$   \\
  $T_0\,^{a}$ (HJD)  & $2454751.536 \pm 0.012$  &  $2454753.95 \pm 0.39$ \\
  $e$                              & $-$  &  $0.031 \pm 0.014$  \\
  $\omega$ (deg)                   & $-$ & $ 251 \pm 26 $ \\
  $K$ (m\,s$^{-1}$)             & $12.72 \pm 0.18$  & $12.65 \pm 0.18 $ \\
  $M_p \sin i$ ($M_{\oplus}$)             & $15.96 $ & $ 15.86 $ \\
  $a$ (AU)                         & $0.041$  &  $ 0.041 $ \\
  \enddata
  \tablenotetext{a}{Predicted mid-transit time.}
\end{deluxetable}

\cite{For11} performed a transit search for GJ 581e using photometry from the 2.5 m Isaac Newton Telescope that was obtained during one transit window between -1 and 2.3 $\sigma$ from the predicted mid-transit time (a total of 6.5 hours) for a circular orbit. They have found no evidence of a transit. In this paper, we present two sets of nearly contiguous MOST space-based photometry of GJ 581, obtained in 2007 and 2009 (section 2). The 2009 data set covers almost four full orbital cycles of GJ 581 e (section 3.1) and thus allows us to perform a more complete transit search using the ephemeris for the circular orbital solution. In addition, we are able to search for transits during the predicted transit window\footnote{The transit window is the time span during which a transit is predicted to occur, calculated from the uncertainties on the orbital period and those on the predicted mid-transit time.} corresponding to the eccentric orbital model. We quantify our sensitivity to shallow transits via Monte Carlo tests (section 3.2). In sections 3.3 and 3.4 we place limits on the size and mass of GJ 581e and discuss the implications. We use the MOST photometry to search for transits of GJ 581b within its circular and Keplerian orbit-predicted transit windows in section 4. In section 5, we briefly address the cases of GJ 581c and GJ 581d. Finally, we combine this data set with a longer light curve obtained in 2007 to carry out a stellar variability analysis for GJ 581 (section 6). We conclude in section 7.

\begin{figure*}[!t]
\begin{center}
\includegraphics[scale=0.35]{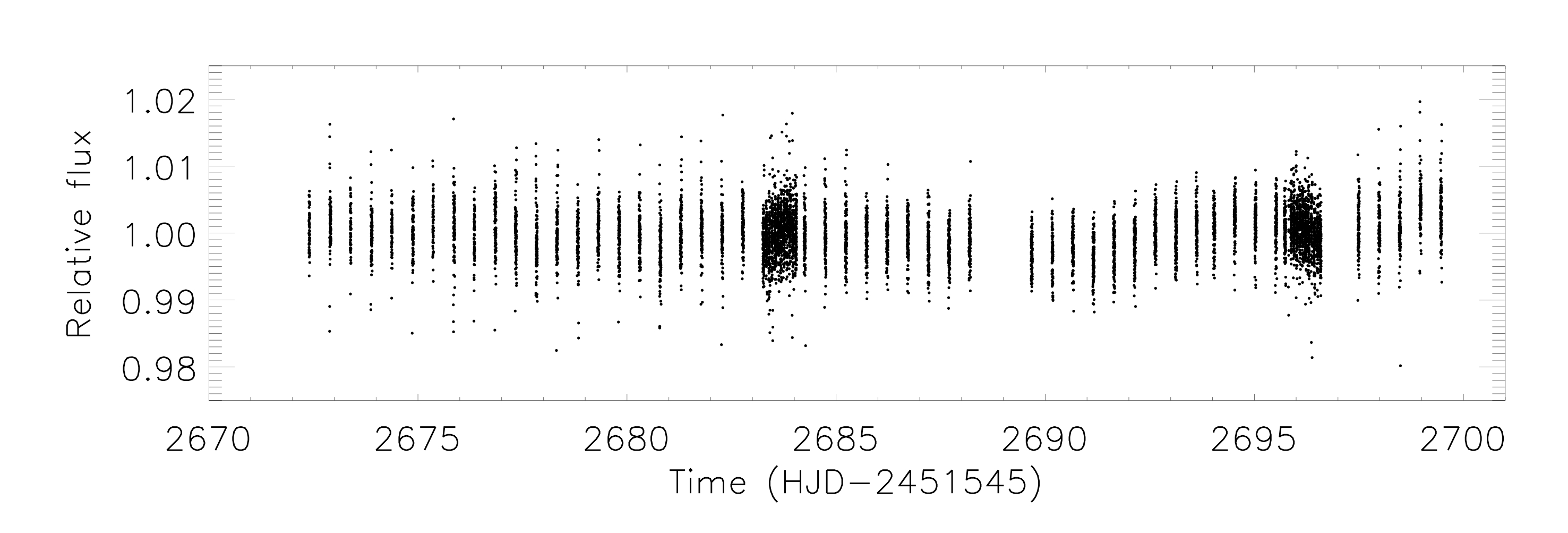}
\caption{Reduced MOST photometry acquired during 2007 April 25 to June 14. The tire-track pattern of the data is due to the fact that MOST alternated between GJ 581 and other stars during these observations. See section 2 of the text for details.}
\end{center}
\end{figure*}

\begin{figure*}[!t]
\begin{center}
\includegraphics[scale=0.35]{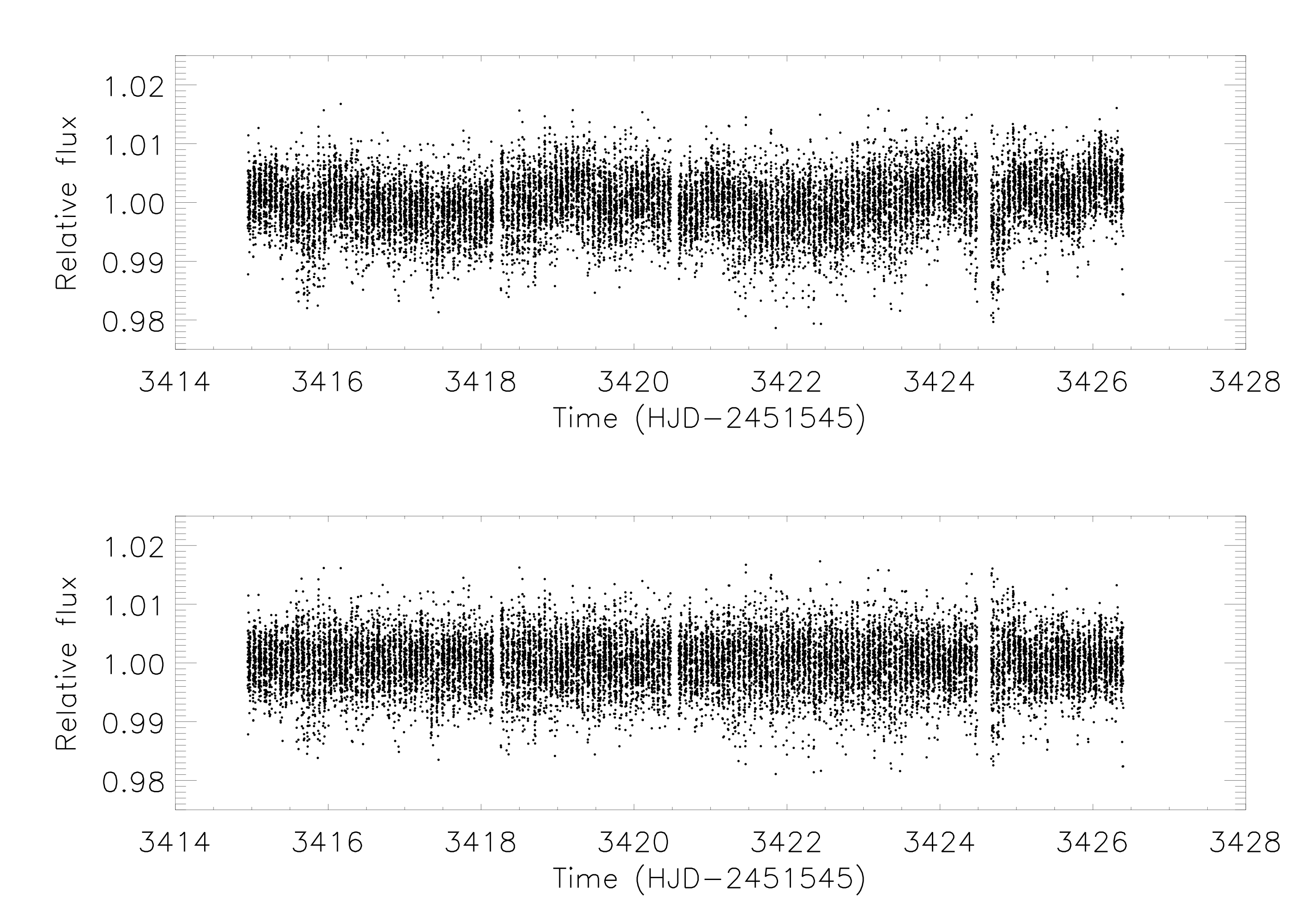}
\caption{{\it Top:} Reduced MOST photometry acquired during 2009 May 8-20. {\it Bottom:} Same, but with the modulation visible in the top panel removed (see section 2 for details). The tire-track pattern of the data is due to the fact that MOST alternated between GJ 581 and another star during these observations. See section 2 of the text for details.}
\end{center}
\end{figure*}

\section{Observations and Data Reduction}

We obtained photometry of Gliese 581 from the MOST satellite during 2007 April 26 to May 24, and during 2009 May 8-20. MOST is a micro-satellite which carries a 15 cm optical telescope feeding a CCD photometer. It is in a Sun-synchronous polar orbit with a period of 101.4 minutes, which allows it to monitor stars in a Continuous Viewing Zone (CVZ) without interruption for up to 8 weeks. The CVZ covers a declination range of -18$^\circ < \delta <$ +36$^\circ$. Stars brighter than V $\sim$ 6 are observed using a Fabry microlens to project onto the CCD an image of the telescope pupil illuminated by the target. Fainter stars are observed in Direct Imaging mode, in which the defocused images of the stars are projected onto the CCD.

In 2007, MOST observed GJ 581 for 50 days (Figure 1), but alternated between this star and other targets during this period. Two stretches of time during which the sampling cadence was significantly higher than for most of the run can be seen in Figure 1. The intention of these two more concentrated ``windows" of coverage was to search for transits of the super-Earth GJ 581c. Unfortunately, due to a mistranslation of the radial velocity ephemeris, these high-cadence sections of the 2007 light curve do not overlap with predicted transit windows for this planet, nor with any predicted mid-transit times for the other known planets in the system. For this reason, and because the sampling outside of these windows is not sufficiently contiguous, the 2007 photometry was not effective for a transit search. Instead, it was combined with the 2009 data to assess the variability of the host star (see section 6). In 2009, GJ 581 was observed contiguously during approximately half of every MOST orbit (54 out of every 101.4 minutes), for nearly 12 days (Figure 2). The alternating target was monitored during the remaining 47.4 minutes. Both the 2007 and 2009 observations were acquired in Direct Imaging mode, with individual exposure times of 3.0 s. Sets of 8 individual frames were stacked on board the satellite before being downloaded to Earth, leading to a total integration time of 24 s per data point. There is also a modest readout overhead of 3.2 seconds, resulting in a sampling rate of 27.2 s (outside of the interruptions for the alternating target). 

The raw data were reduced using aperture photometry. The reduction pipeline (described in detail in \citealt{Row08}) corrects for cosmic ray hits and stray light from scattered Earthshine, which varies with the period of the satellite. The 1-day period variation (due to the Sun-synchronous orbit of MOST) with an amplitude of 1.3 mmag was also filtered from the data. 

In addition, for the 2009 data set three short sections of the light curve (at around 3418.2, 3420.5 and 3424.6 days in Figure 2) affected by high levels of stray light were conservatively excised. After these steps, a modulation remains in the light curve (Figure 2, top panel). A possible origin of this modulation is discussed in section 6. This trend was modelled with a cubic spline interpolated from the data binned every 500 minutes, and removed from the light curve before the transit search and transit-injection tests. The bin size was chosen to be large enough not to interfere with any transits of GJ 581 e (predicted edge-on transit duration = 1.2 hours) and small enough to remove the slowly varying trend. The final 2007 and 2009 time series contain 7667 and 17312 points, respectively.

\section{GJ581\lowercase{e}}

\subsection{Transit Ephemeris and Predicted Characteristics}

We use the ephemerides published in \cite{For11} to search for transits of GJ 581e. The authors derived best-fitting orbital models for planets b, c, d and e in two ways: by assuming circular orbits (where the eccentricity is fixed to 0.0 for each planet) and by fitting fully Keplerian orbits (where the eccentricity is allowed to float). While the values of the orbital period for GJ 581e in the two cases are nearly identical, an eccentricity of 0.32 $\pm$ 0.09 is found for the Keplerian fit (see Table 1 for the \citealt{For11} orbital parameters of GJ 581e). In addition, $M_{P} \sin i$ is slightly larger for the Keplerian orbit fit (1.95 $M_{\oplus}$) than for the circular orbit model (1.84 $M_{\oplus}$), though no uncertainties are reported for these values. The predicted mid-transit times ($T0$) and associated uncertainties also differ significantly between the two cases, leading to different phase locations and sizes for the transit windows. The coverage of the MOST photometry is adequate for monitoring the transit windows corresponding to both solutions.

For the circular orbit transit search, the photometry were folded on $T0=$2454969.489. The 1$\sigma$ uncertainty on this value is 0.071 days, so the size of the 3$\sigma$ transit window is 11.48 hours (including a predicted transit duration of 1.2 hours). In the Keplerian orbit case, $T0=$2454970.77 was used to fold the time series. The associated 1$\sigma$ uncertainty is 0.14 days, leading to a 3$\sigma$ transit window of 22.03 hours (including a predicted transit duration of 1.6 hours).

We estimate radii for GJ 581e for each of three theoretical compositions: hydrogen, water ice and silicate. We use the models of \cite{Sea07} and the planetary mass obtained by assuming an impact parameter ($b$) of 0.0 ($i=90^{\circ}$). The derived radii are 3.70, 1.68 and 1.25 $R_{\oplus}$, for a homogeneous hydrogen, water ice and silicate planet, respectively.

\subsection{Monte Carlo simulations}

Throughout this section we only use the 2009 data set because the sampling cadence of the 2007 photometry is too low for a transit search of a super-Earth planet.

Our approach is inspired by the methods used in \cite{Cro07a}, \cite{Cro07b} and \cite{Bal10}. However, our analysis differs in that we search for transits of a RV-detected planet with a well-known period rather than additional planets in known transiting systems. Indeed, the orbital period of GJ 581 e is known to within 19 s \citep{For11}. As a consequence, for our Monte Carlo simulations and transit search we vary the planetary radius ($R_{p}$), orbital inclination ($i$) and orbital phase at which the transit may occur ($0<\phi<1$), but not the orbital period. Carrying out simulations for values of $i$ other than 90$^{\circ}$ allows us to quantify the effect of the sampling cadence of the MOST observations on our sensitivity to transits of various durations.

To quantify our detection limits for the Gliese 581 photometry, we injected simulated limb-darkened transits (using the models of \citealt{Man02}) in the 2009 light curve and attempted to recover them by fitting a box-shaped transit model and minimizing $\chi^2$. For the simulated transits, we interpolated limb darkening coefficients ($c_{1}=$1.600, $c_{2}=$-1.491, $c_{3}=$1.209, $c_{4}=$-0.398) from tables generated for the MOST bandpass (A. Prsa, private communication) and based on the Kurucz models \citep{Cas04}, assuming a stellar effective temperature ($T_{eff}$) of 3500 K, surface gravity (log $g$) of 5.0 and metallicity ($[Fe/H]$) of -0.15. The actual values for GJ 581 are $T_{eff}=3498 \pm 56$ K, log $g = 4.96 \pm 0.08$ and $[Fe/H] = -0.135$ \citep{Bra11}. We used a grid evenly spaced in radius ($R_{P-inj}$) and orbital inclination ($i_{inj}$). The grid consists of 30 $R_{P-inj}$ values ranging from 1 to 4 $R_{\oplus}$ and 4 $i_{inj}$ values ranging from 87.5 to 90$^{\circ}$ ($0.0 < b < 0.9$). For each combination of radius and inclination, we insert transits at 100 randomly distributed phases ($\phi_{inj}$) spanning the 3$\sigma$ transit window for the circular ephemeris. Thus we search a total of 12000 light curves.

We sample the $\chi^2$ space at 90 values of the planetary radius (equally spaced from 0.85 to 4.5 $R_{\oplus}$ ), 6 values of the orbital inclination (equally spaced from 87.2 to 90$^{\circ}$) and between 130 and 2000 values of the orbital phase (equally spaced within the 3$\sigma$ transit window). For each model, we compute the change in $\chi^{2}$ relative to the constant flux hypothesis. The model with the largest value of $\Delta\chi^{2}\%=100\%$($\chi^{2}_{t}-\chi^{2}_{c}$)$/\chi^{2}_{c}$ (where $\chi^{2}_{t}$ and $\chi^{2}_{c}$ correspond to the transit and constant flux model, respectively) becomes the best-fit solution for a particular light curve. We call the values of $R_{p}$, $i$ and $\phi$ associated with the best-fit solution $R_{rec}$, $i_{rec}$ and $\phi_{rec}$, respectively. Since the goal is to identify a statistically significant transit rather than to precisely determine its parameters, we consider a transit as successfully detected if the following two conditions are satisfied: $|(R_{rec}/R_{inj})|-1 < 20 \%$ and $|\phi_{rec}-\phi_{inj}| < 0.01$ (corresponding to $\sim$ 45 min). $R_{inj}$ and $\phi_{inj}$ are the planetary radius and orbital phase of the injected transit signal. The results of our Monte Carlo simulation indicate that we do not always accurately recover the impact parameter of the injected signal. This is not surprising for such shallow and short (1.2 hours for $b=0$) transits. As a consequence, we do not use an orbital inclination condition as a criterion for detection.

For $b=0$ ($i = 90^{\circ}$), we recover the transits corresponding to a planetary radius $\geq$1.42 $R_{\oplus}$ at least 68$\%$ of the time. We recover injected transits for $R_{p}\geq1.62 R_{\oplus}$ and for $R_{p}\geq1.93 R_{\oplus}$ at least 95$\%$ and 99$\%$ of the time, respectively. For larger values of $b$ these confidence limits tend to shift to larger radii, though only mildly so for $b \leq$ 0.6. This is because the transit duration is shorter and there are less in-transit data points, making the signal more difficult to detect. Further, for larger values of $b$ the planet would cross nearer the limb of the star, thus blocking out less light and leading to a shallower transit than in the case of $b=0$. In light of these two arguments, we note that while transits with $R_{P}\leq4 R_{\oplus}$ and $b=0.9$ do not meet our detection criteria with sufficient statistical confidence, deeper transits do. Our detection limits as a function of $b$ and $R_{p}$ are shown in Figure 3, in the form of 68$\%$, 95$\%$ and 99$\%$ confidence contours. A value of $b=0$ corresponds to an edge-on transit, while $b=1$ represents a grazing transit.

\begin{figure}[!h]
\begin{flushleft}
\includegraphics[scale=0.25]{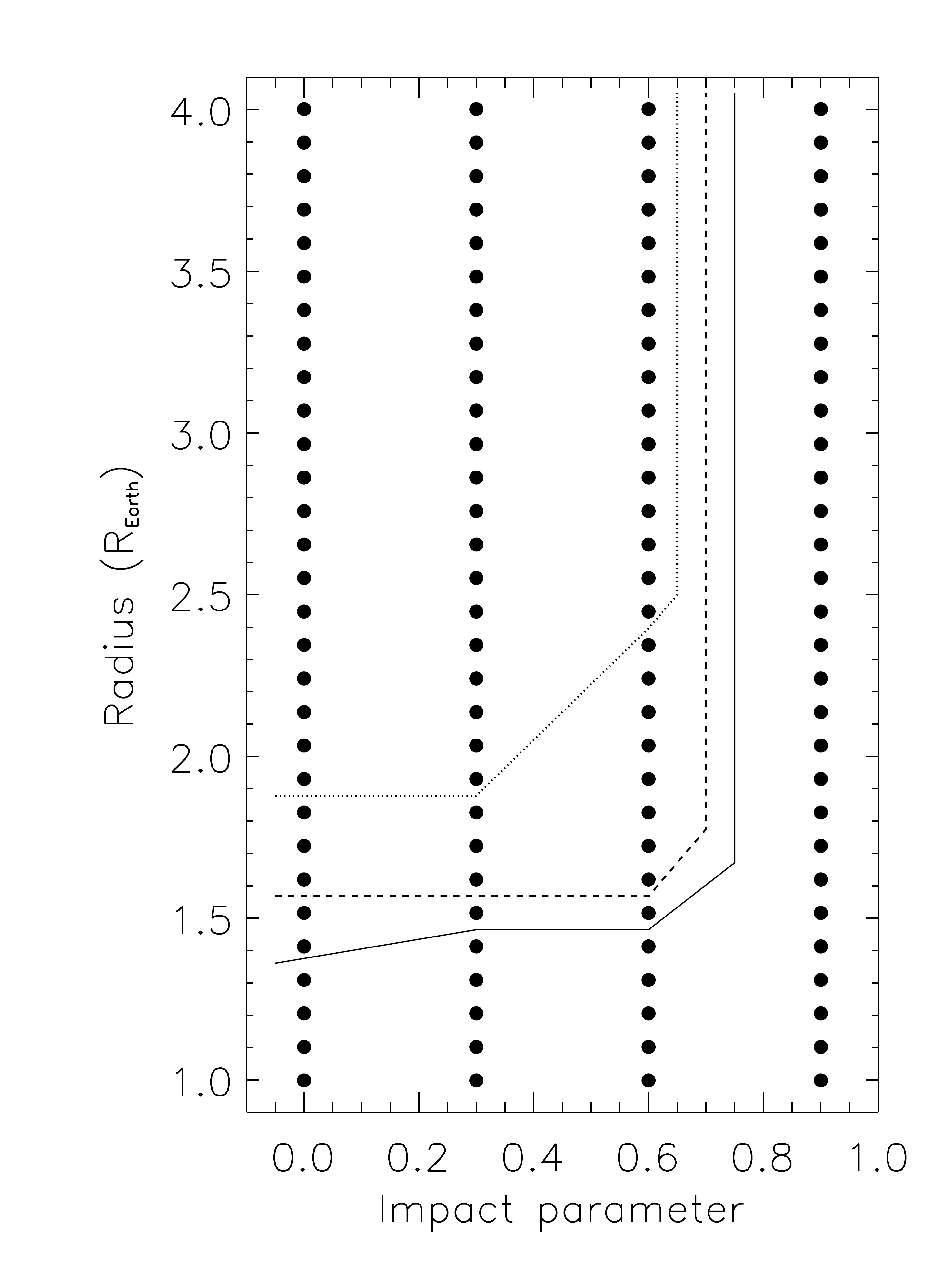}
\caption{Detection limits for injected transits as a function of planetary radius ($R_{P-inj}$) and impact parameter ($b_{inj}$), for GJ 581e. The solid points correspond to the values of $R_{P-inj}$ and $b_{inj}$ at which we tested our sensitivity. The solid, dashed and dotted lines indicate the 68$\%$, 95$\%$ and 99$\%$ confidence contours, respectively. }
\end{flushleft}
\end{figure}

%\section{Results and Implications}

\subsection{Transit Search}

We perform the transit search described above on the 2009 light curve (within the 3$\sigma$ transit window corresponding to the circular orbit solution). We do not find any transit candidates for a planet with $R_{p}\geq$1.42 $R_{\oplus}$, which is our 1$\sigma$ detection limit (for $b=0$). 

Our best transit candidate (for which $\Delta\chi^{2}\%=0.54 \%$ was largest) has the following properties: R$_{p}=0.912 R_{\oplus}$, $b=0.40$ and $\phi=0.058$, and therefore it occurs near the positive edge of the 3$\sigma$ window (the predicted mid-transit time corresponds to $\phi=0.0$). This signal is significantly below our detection limit and over five times smaller than the rms of the phased data within the 3$\sigma$ transit window. Such a planet would have a density ($\rho$) of 13.4 g cm$^{-3}$, an unrealistic value as it is greater than the maximum iron fraction limit for a rocky planet \citep{Mar10}. For these reasons, we consider this candidate very unlikely to be a transit signal. The phased observations spanning the 3$\sigma$ transit window and the best-fit model for this candidate are shown in Figure 4.

Using the HIRES-based results from \cite{Vog10}, \cite{And10} find a possible discrepancy between the residuals of a circular orbit solution for the four planets, and the expected Gaussian distributed residuals if the best-fit model was the true model. They suggest this discrepancy may indicate that the planetary orbits are eccentric. Further, when fitting a Keplerian model to the HARPS RV data (see section 3.1), \cite{For11} obtain significant values for the eccentricity of planets d and e, and different mid-transit times for all planets than in the circular model case. Motivated by these result, we also carry out a transit search for GJ 581e within the 3$\sigma$ transit window computed from the orbital parameters associated with the Keplerian model. The rms=0.00425 (in units of relative flux) of the phased photometry within this window is marginally smaller than that of the circular solution transit window (rms=0.00431). Thus our sensitivity to transits should be very similar or marginally better. We find no transit candidate near or above the detection limits reported in Figure 3. 

The best candidate ($\Delta\chi^{2}\%=0.92 \%$) in this case represents a planet with R$_{p}=1.19 R_{\oplus}$, $b=0.79$ and $\phi=-0.006$, very close to the predicted mid-transit time. With a mass of 1.95 M$_{\oplus}$ \citep{For11}, this planet would have $\rho=$6.41 g cm$^{-3}$ which would make it slightly denser than 55 Cnc e and suggest a silicate composition model \citep{Sea07}. Nevertheless, the statistical significance of this signal is too low relative to our detection limits to make it a likely candidate. The phased observations showing this candidate can be seen in Figure 5.

\begin{figure}[!h]
\begin{flushleft}
\includegraphics[scale=0.215]{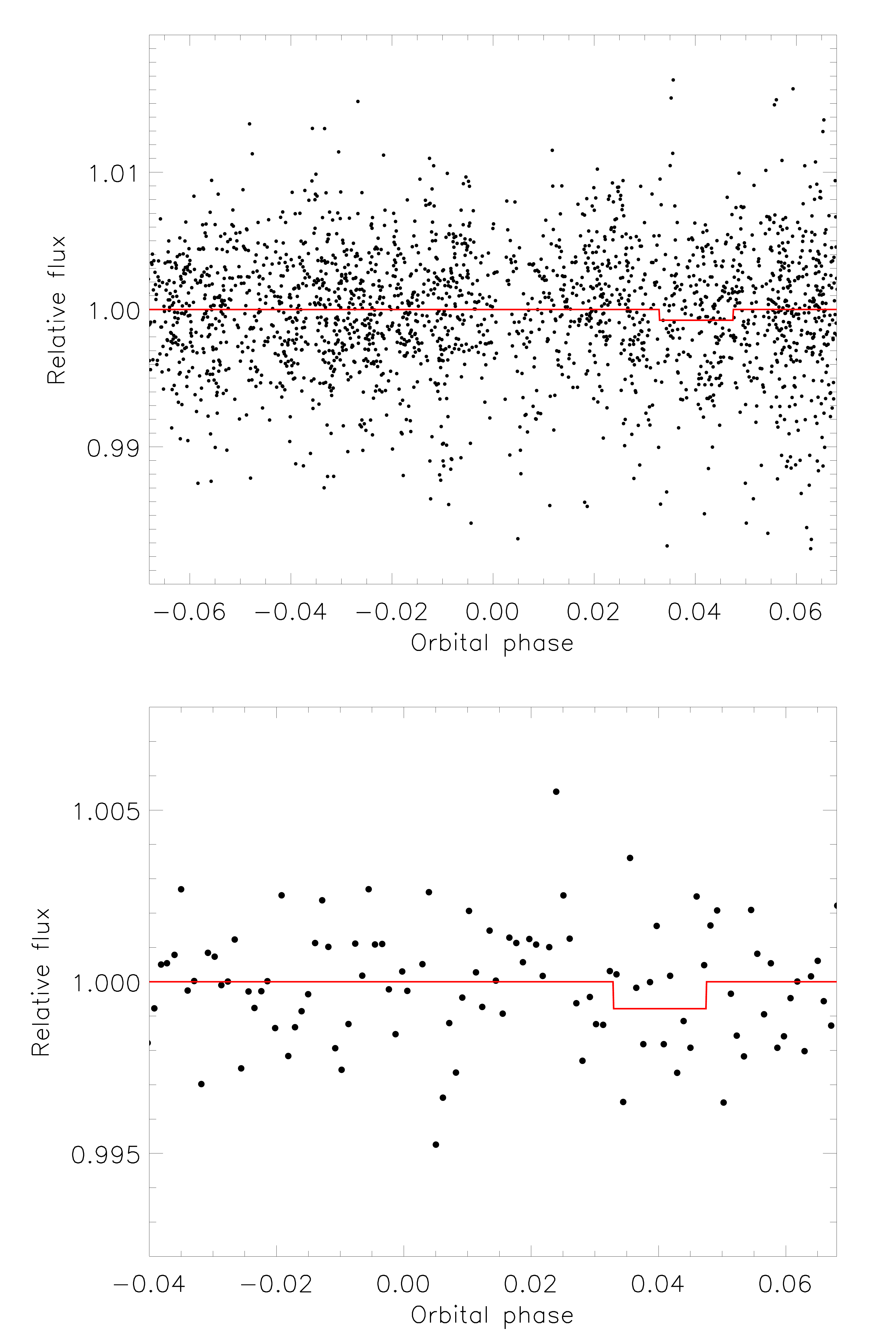}
\caption{The best candidate transit signal for the circular model ephemeris of GJ 581e. Phase 0.0 corresponds to the predicted mid-transit time. {\it Top:} Unbinned photometry folded at the orbital period of 3.15 days. The orbital phase range of the plot is equivalent to the 3$\sigma$ transit window. {\it Bottom:} Zoom-in of the folded photometry, binned every 0.001 orbital phase. In both panels, the best-fit box-shaped transit model is shown in red.}
\end{flushleft}
\end{figure}

\begin{figure}[!h]
\begin{flushleft}
\includegraphics[scale=0.215]{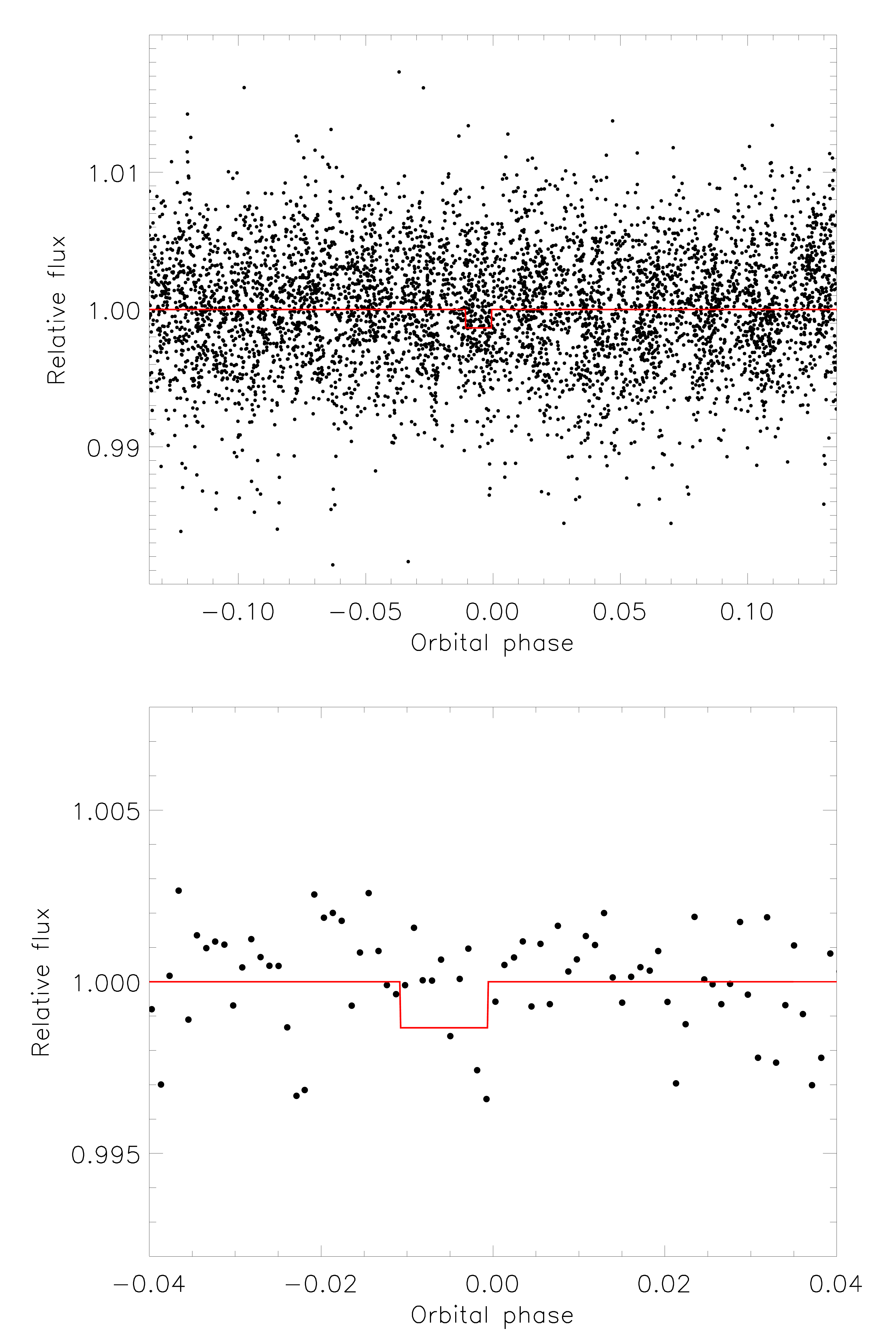}
\caption{The best candidate transit signal for the Keplerian model ephemeris of GJ 581e. Phase 0.0 corresponds to the predicted mid-transit time. {\it Top:} Unbinned photometry folded at the orbital period of 3.15 days. The orbital phase range of the plot is equivalent to the 3$\sigma$ transit window. {\it Bottom:} Zoom-in of the folded photometry, binned every 0.001 orbital phase. In both panels, the best-fit box-shaped transit model is shown in red.}
\end{flushleft}
\end{figure}

\subsection{Constraints and Implications}

Even when a super-Earth transits and its radius can be measured, it is still difficult to gain insight into its detailed composition. Nonetheless, useful limits can be placed on its interior structure when both its mass and radius are available \citep{Rog10}. Unfortunately, we do not detect a significant transit signal for GJ 581e. If the planet transits, then we can use the limits determined in section 3.2 to constrain its size as follows.

To 1$\sigma$ confidence, we exclude pure hydrogen, hydrogen/helium, pure water ice and 50$\%$ water ice compositions for detectable ($b\leq0.6$) transiting orbital configurations. To 2$\sigma$ confidence, we rule out compositions less dense than 75$\%$ water for $b\leq0.6$. We exclude hydrogen and hydrogen/helium compositions with 3$\sigma$ confidence. Quantitatively, for $M_{P} = 1.84 M_{\oplus}$ (based on the circular orbit solution), we rule out densities lower than 3.54 g cm$^{-3}$ , 2.39 g cm$^{-3}$ and 1.41 g cm$^{-3}$ with 1$\sigma$, 2$\sigma$ and 3$\sigma$ confidence, respectively.

There are so far no transiting planets with a measured mass smaller than 2 $M_{\oplus}$. Kepler-11f comes close, at 2.3 $M_{\oplus}$ (determined from dynamical fits of the Kepler-11 system) and a radius of 2.6 $R_{\oplus}$ \citep{Lis11}. GJ 1214b orbits a star much closer in spectral type (M4.5) to GJ 581, but has a mass of 6.4 $M_{\oplus}$ and a radius of 2.7 $R_{\oplus}$ \citep{Cha09}. These two super-Earths are relatively low-density planets, with envelopes of water \citep{Ber12} and/or hydrogen \citep{Lis11}. If GJ 581e transits and has a similar density, then we would have detected the transit in the MOST photometry. However, if a transiting GJ 581e is more similar to the denser 55 Cnc e ($\rho=$5.9 g cm$^{-3}$; \citealt{Win11}), Kepler-10b ($\rho=$8.8 g cm$^{-3}$; \citealt{Bat11}), or Corot-7b ($\rho=$10.4 g cm$^{-3}$, assuming a mass of 7.4$M_{\oplus}$; \citealt{Hat11}), then we would not have detected its transit using our data. Nevertheless, 55 Cnc e, Kepler-10b and Corot-7b receive significantly more radiation due to both their proximity to their host stars and the higher temperatures of those stars, compared to Kepler-11f, GJ 1214b and GJ 581e. Therefore, we expect any atmospheres or water envelopes these highly irradiated planets may have acquired to be mostly or entirely evaporated, thus possibly explaining their higher densities.

\cite{May09} have performed a stability analysis of the GJ 581 system in order to determine constraints on its orbital inclination. Assuming co-planar orbits and allowing for non-zero eccentricity, they found that GJ 581e was ejected for inclinations less than $i = 40^{\circ}$. This lower bound leads to a maximum mass for GJ 581e of 2.86 $M_{\oplus}$ based on the $M_{P} \sin i$ value from the circular solution, and 3.03 $M_{\oplus}$, based on the Keplerian solution. A co-planar orbital configuration with $i > 40^{\circ}$, even if it is not edge-on, should still give rise to observable thermal phase variations of the order of at least 10 ppm in the infrared, dominated by the thermal radiation of planets b (see section 4) and e. A signal of this amplitude is likely within reach of NIRCam on the JWST \citep{Sea09}.

%%%%%%%%%%%%%%%%%%%%%%%%%%%%%%%%%%%%%%%%%%%%%%%%%%%%%%%%%%%%%%

\section{GJ 581\lowercase{b}}

\subsection{Transit Ephemeris and Predicted Characteristics}

We use the \cite{For11} ephemerides for GJ 581b to determine its 3$\sigma$ transit window for both the circular and Keplerian orbit cases, in the same manner we did for GJ 581e. Although the authors find a nearly circular orbit ($e = 0.031 \pm 0.014$) for GJ 581b itself in the Keplerian orbit case, the planet's predicted mid-transit time is nevertheless affected (and shifted relative to the circular case) due to the overall best-fitting solution which does assign significantly greater than 0 eccentricities to planets $e$ and $d$. The minimum mass of GJ 581b is not very different in the circular case ($M_{P}\sin i = 15.96 M_{\oplus}$) compared to the Keplerian case ($M_{P}\sin i = 15.86 M_{\oplus}$), and is very close to that of Neptune ($M_{P} = 17.15 M_{\oplus}$).

We find that \cite{Lop06} have conclusively ruled out transits of GJ 581b within its $\sim$ 2$\sigma$ transit window for the circular orbit ephemeris. They exclude planetary radii greater than $1.5 R_{\oplus}$, a value which, given the minimum mass of the planet, would suggest a highly implausible object composed entirely of iron. However, to our knowledge the transit window arising from the Keplerian fit has not yet been searched. 

For completeness (the MOST observations allow the search to be carried out beyond the 2$\sigma$ window), and to confirm the results of \cite{Lop06}, we perform this analysis for the circular orbit window as well, using $T0=2454966.303$ to fold the photometry. The 1$\sigma$ uncertainty on this value is 0.016 days. The 3$\sigma$ transit window (including a predicted edge-on transit duration of 1.4 hours) is thus 3.7 hours long. For the Keplerian orbit window we use $T0=2454963.34$ to fold the data, and the associated 1$\sigma$ uncertainty of 0.39 days leads to a transit window length of 56.2 hours (or 2.34 days).

\subsection{Transit Search and Constraints}

The folded photometry within the 3$\sigma$ transit window for the circular and Keplerian orbit cases are shown in Figures 6 and 7, respectively. We considered and overplotted transit models corresponding to three possible planetary radius values. The dashed line represents the minimum radius for a planet of mass equal to the minimum mass of GJ 581b with the maximum iron core mass fraction attainable from collision-induced mantle stripping for a planet ($R_P = 1.9 R_{\oplus}$; \citealt{Mar10}). A more realistic transit depth is indicated by the dotted line, which corresponds to a silicate planet ($R_P = 2.3 R_{\oplus}$). Finally, the solid line represents a planet with a density equivalent to that of Neptune ($R_P = 3.8 R_{\oplus}$).

\begin{figure}[!h]
\begin{flushleft}
\includegraphics[scale=0.215]{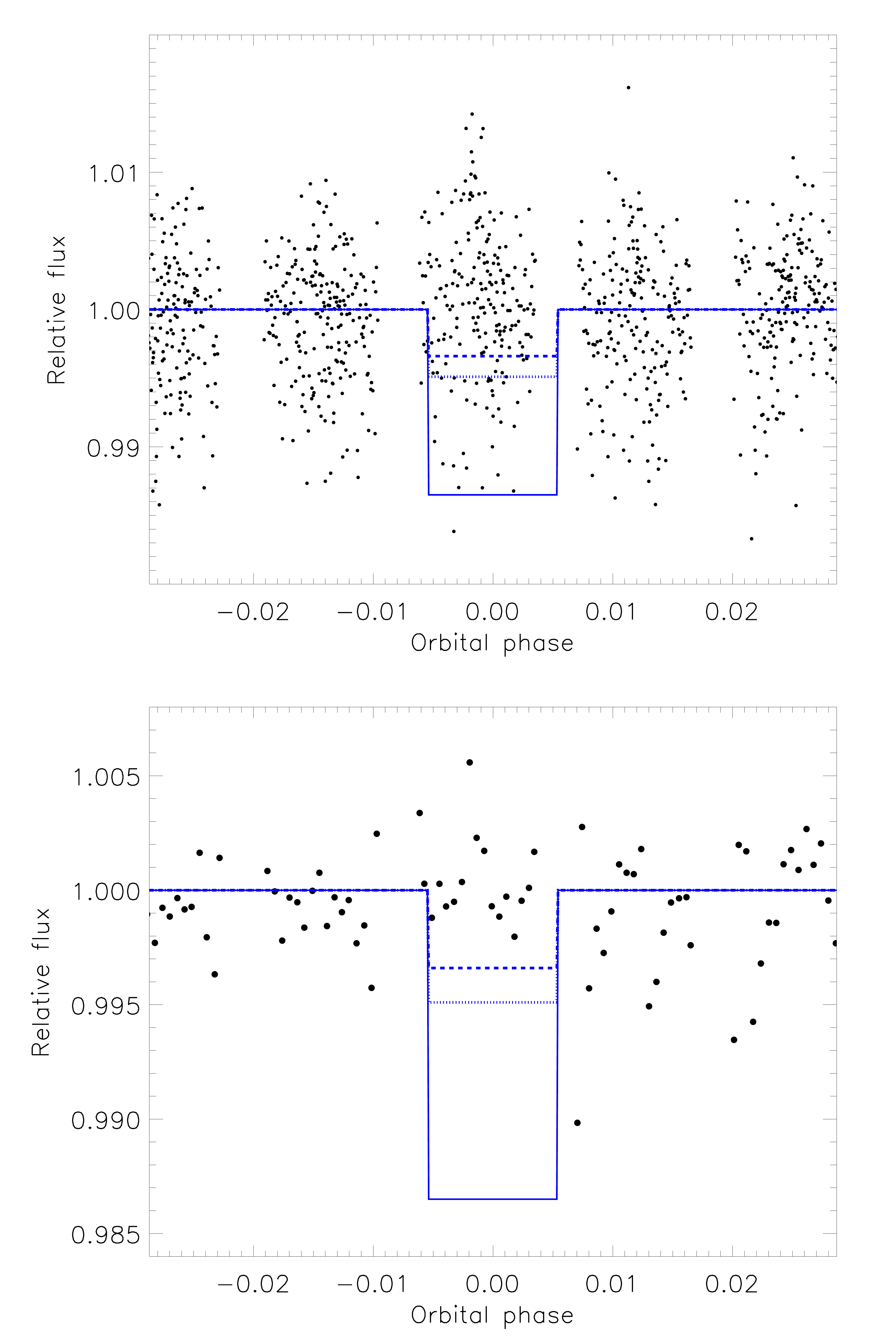}
\caption{The 2009 photometry, folded at the period of GJ 581b (5.37 days). The orbital phase range of both top and bottom plots is equivalent to the 3$\sigma$ transit window based on the circular orbit ephemeris. Phase 0.0 corresponds to the predicted mid-transit time. Box-shaped predicted transit signatures for three planetary radius values are overplotted in blue. The dashed, dotted and solid lines represent $R_P = 1.9, 2.3$ and $3.8 R_{\oplus}$, respectively. {\it Top:} Unbinned photometry.  {\it Bottom:} Vertical zoom-in of the folded photometry, binned every 0.0006 orbital phase.}
\end{flushleft}
\end{figure}

\begin{figure*}[!t]
\begin{center}
\includegraphics[scale=0.35]{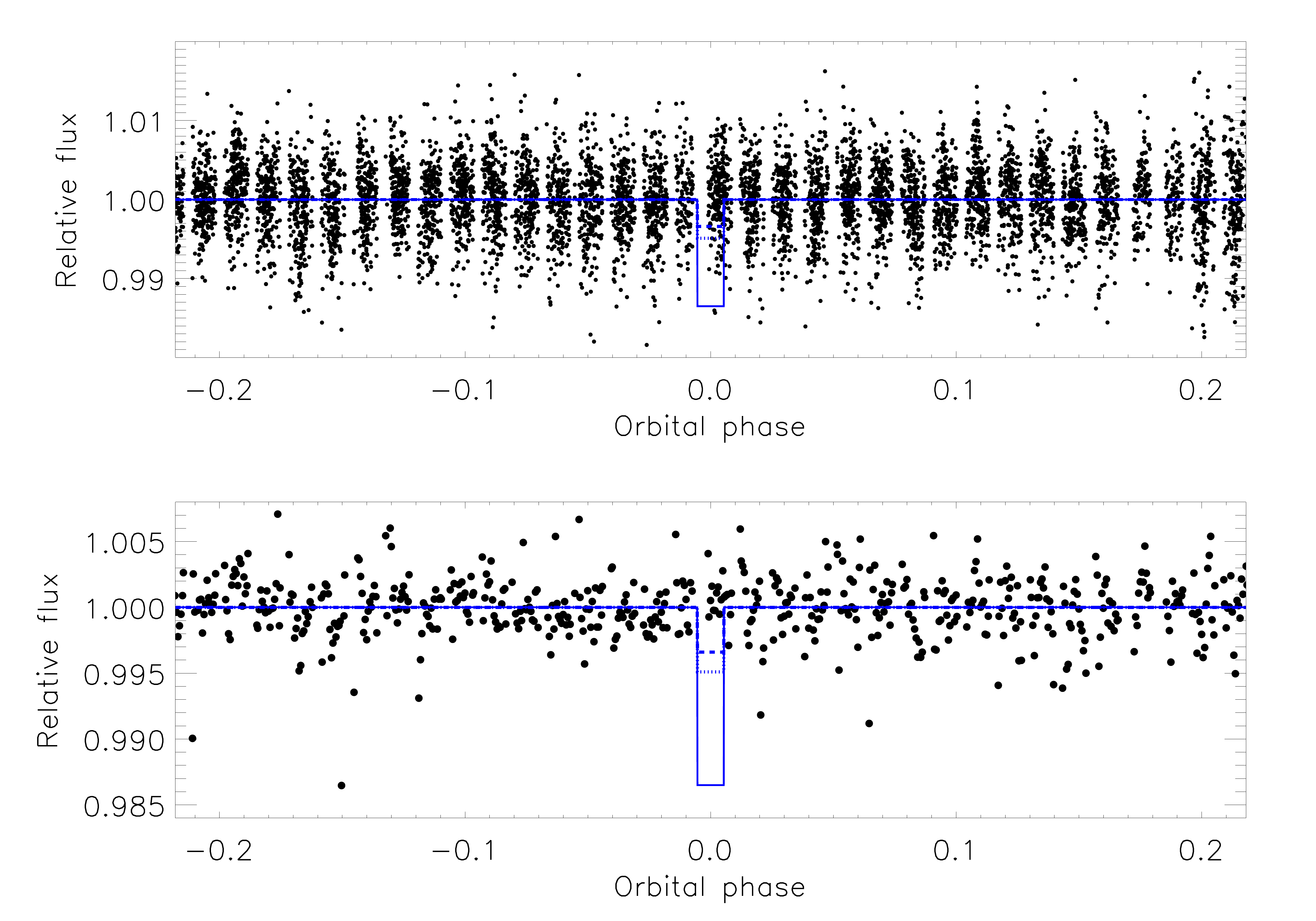}
\caption{The 2009 photometry, folded at the period of GJ 581b (5.37 days). The orbital phase range of both top and bottom plots is equivalent to the 3$\sigma$ transit window based on the Keplerian orbit ephemeris. Phase 0.0 corresponds to the predicted mid-transit time. Box-shaped predicted transit signatures for three planetary radius values are overplotted in blue. The dashed, dotted and solid lines represent $R_P = 1.9, 2.3$ and $3.8 R_{\oplus}$, respectively. {\it Top:} Unbinned photometry.  {\it Bottom:} Vertical zoom-in of the folded photometry, binned every 0.0006 orbital phase.}
\end{center}
\end{figure*}

The phase coverage for GJ 581b is incomplete even when the $\sim$2 orbital cycles spanned by the observations are phase-folded because of the sampling cadence of the MOST photometry of this system (see section 2). This sets limits on the duration of a transit detectable with the MOST data, depending on where it would occur within the window.

Thus, by visual inspection of the phased light curve within the predicted transit window, we can rule out transits corresponding to radii associated with any plausible composition ($R_P \geq 2.3 R_{\oplus}$) for GJ 581b, but only for impact parameter $b\lesssim0.4$ (corresponding to a transit duration greater than $\sim 90\%$ of that of an edge-on transit). These limits apply to both the circular and Keplerian orbit cases. It is difficult to set more precise (and applicable to the entire transit window) limits on the duration of a detectable transit because the sampling cadence and phase coverage of the phased photometry are uneven throughout the transit window.

For the circular orbit window -- based on the results of \cite{Lop06} and our own -- we conclude that transits are unlikely to occur. For the Keplerian case, given our observations, transits may still occur for $0.4\lesssim b \leq 1$. 

The lower limits on the mass of GJ 581b thus do not change. The dynamical simulations performed by \cite{May09} (under the assumption of stable and co-planar orbits for planets a to d) lead to an upper limit of 24.83 $M_{\oplus}$ for the circular orbit case and 24.67 $M_{\oplus}$ for the Keplerian orbit case.

\section{GJ 581\lowercase{c} and GJ 581\lowercase{d}}

The third and fourth known planets in the GJ 581 system have orbital periods of 12.92 and 66.6 days, respectively. They are both super-Earth candidates, with a minimum mass of 5.4 $M_{\oplus}$ for GJ 581c and 5.3 $M_{\oplus}$ for GJ 581d, in the circular orbit case. Their long orbital periods result in low transit probabilities, but having photometry spanning several days to several weeks motivated us to consider whether the observations may include the predicted mid-transit times of these two planets. 

Unfortunately, no GJ 581 d predicted transit times occur within the 2007 or 2009 data sets, for either the Keplerian or circular orbit ephemerides. No GJ 581c transit times are predicted within the high-cadence sections of the 2007 light curve (visible in Figure 1), but the 2009 data set does include one value of $T0$ for each of the two orbital fits. For the circular case, $T0=2454970.93$ with a 1$\sigma$ uncertainty of 0.15 days. For the Keplerian case, $T0=2454969.7$ with a 1$\sigma$ uncertainty of 1.6 days. 

The main reason we do not carry out a transit search beyond visual inspection of the light curve is that the MOST photometry only covers one transit window for GJ 581c. Most of the possible radius values for this planet are too small to produce a transit signal that would be believable with a single detection given the precision of our photometry. The relatively long transit windows, especially for the Keplerian orbit case, would further decrease the significance of such a signal. We do note that no transit feature with depth larger than the rms of the light curve ($\sim 0.005$) is visible in the photometry. This limit corresponds to a radius of 2.3 $R_{\oplus}$ in the case of an edge-on transit. Given the relatively low mass of GJ 581c (5.4 $M_{\oplus}$) this is not a particularly constraining limit on the radius of this planet. If GJ 581cÊtransits the host star, this limit still allows a range of densities which include water/helium/hydrogen planetary compositions.

%%%%%%%%%%%%%%%%%%%%%%%%%%%%%%%%%%%%%%%%%%%%%%%%%%%%%%%%%%%%%%%

\section{Stellar Variability}

The brightness of GJ 581 is stable to less than 1$\%$ over the four weeks during which MOST observed the star in 2007. The 11.5-day data set obtained in 2009 supports this conclusion. This level of stability agrees with the finding that the X-ray brightness of GJ 581 lies below ROSAT's detection threshold and supports an advanced age ($\gtrsim$ 7 Gyr) for the star, as suggested by \cite{Sel07}. Thus, it constitutes a factor in favour of the potential development of life on any planet residing within the habitable zone of the system, and allows a sufficiently long period of time for its occurrence \citep{Sca07}. 

We searched the MOST photometry for evidence of variations indicative of the stellar rotation period. \cite{Vog10} report a value of 94.2 days, based on APT photometric observations acquired with the 0.36 m Tennessee State University telescope and spanning 530 nights. They find a semi-amplitude of 3 mmag (0.0028 in units of flux) for this variation. Though it covers only 31$\%$ of the reported stellar rotation period, the MOST 2007 data is contemporaneous with a subset of the 0.36 m photometric observations ranging from 2454217.3 to 2454244.5 (rotational phase 0.84 to 1.13 according to section 4 and Figure 1 of \cite{Vog10}). 
The MOST 2007 photometry shows a slight long-term curvature, with a minimum at approximately HJD 2454236. This corresponds to phase 1.05 in Figure 1 of \citealt{Vog10}. In their figure, the minimum of their folded photometry can also be observed near the same phase. However, our data are insufficient to definitively establish whether they support a rotation period of 94.2 days.

In the 2009 data set we observe a modulation in the light curve with a period of about 5 days and semi-amplitude of $~$1.5 mmag (or 0.14$\%$ in flux units; see Figure 2), which was removed prior to the transit injection tests and transit search as described in section 2. The amplitude spectrum shows a strong peak at a frequency of 0.1790 $\pm$ 0.0016 cycles/day (period of 5.586 $\pm$ 0.051 days), as can be seen in Figure 9. This variation is especially interesting as it has a period comparable to the orbital period of GJ 581b ($P = 5.36865 \pm 0.00009$ days). We generate an amplitude spectrum of the 2007 photometry (see Figure 8) to verify whether the same signal is present. A peak of much lower significance is visible at a frequency corresponding to a period of 5.365 $\pm$ 0.074 days. In Figure 10, we show the 2007 and 2009 photometry phased to the planet's orbital period, separately and together. While the 2007 data is not persuasive, the 2009 photometry shows a coherent signal. This signal persists when the two data sets are combined, and while this is at least in part due to the greater number of observations in the 2009 photometry which dominate the combined data, its persistence is arguably encouraging.

\begin{figure}[!h]
\begin{flushleft}
\includegraphics[scale=0.215]{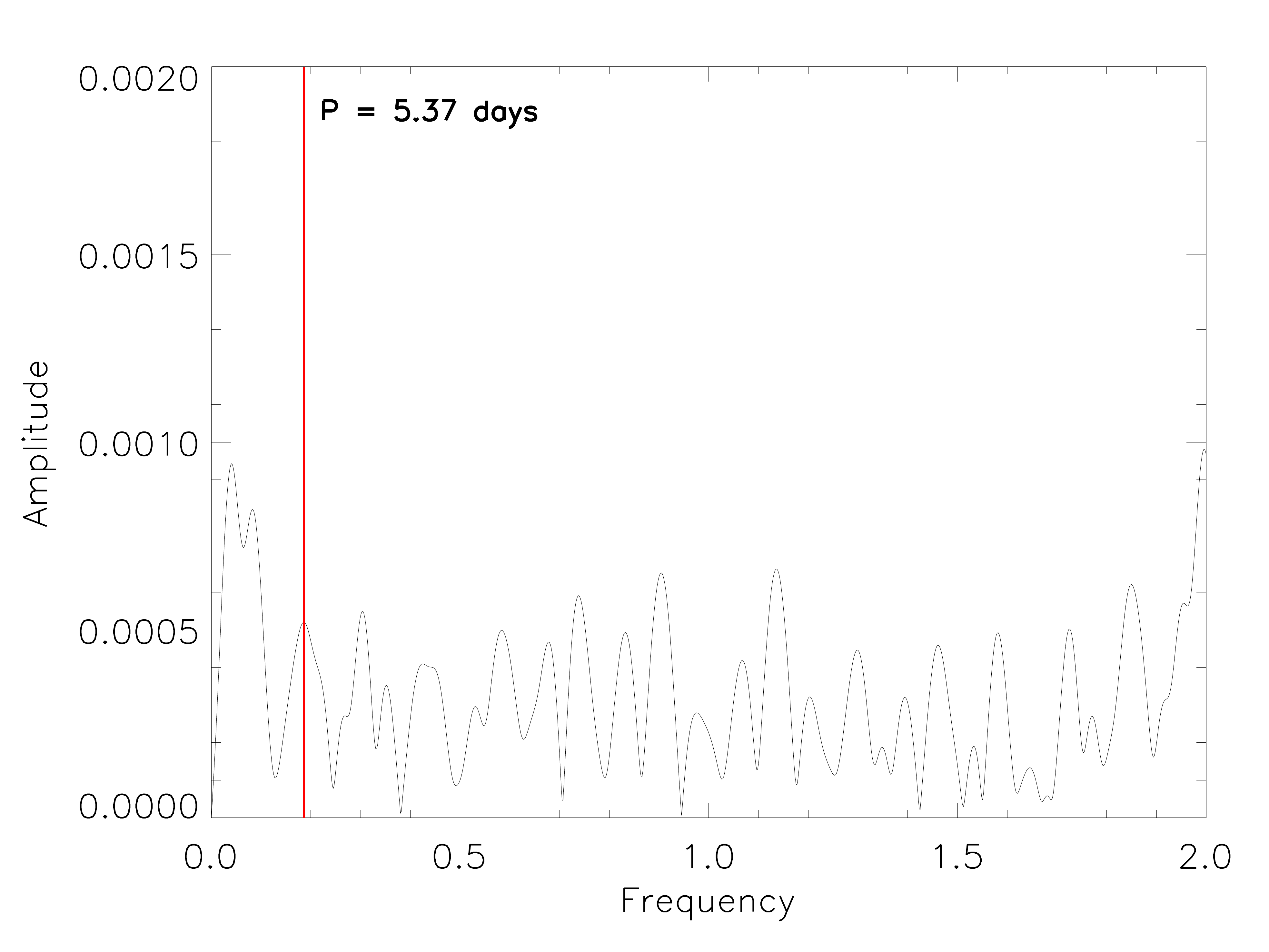}
\caption{Amplitude spectrum (in units of relative flux) for the 2007 MOST photometry. The red line indicates the frequency corresponding to the orbital period of GJ 581b ($P=5.37$).}
\end{flushleft}
\end{figure}

\begin{figure}[!h]
\begin{flushleft}
\includegraphics[scale=0.215]{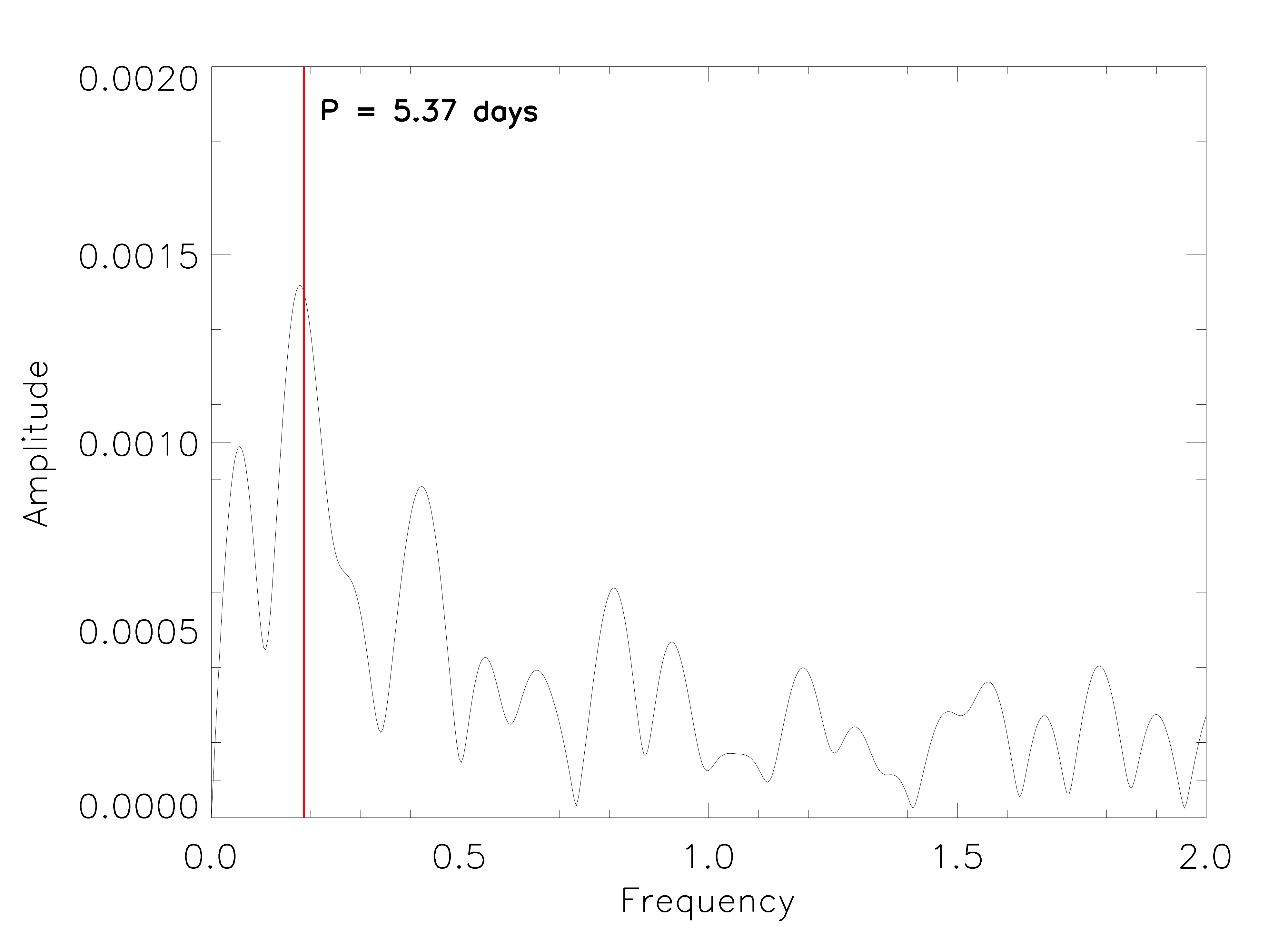}
\caption{Amplitude spectrum (in units of relative flux) for the 2009 MOST photometry. The red line indicates the frequency corresponding to the orbital period of GJ 581b ($P=5.37$).}
\end{flushleft}
\end{figure}

Regardless of whether this variation persists beyond the duration of our 2009 observations, its period is too short to correspond to the stellar rotation period, as pointed out by \cite{May09}. Considering the approximate match between the orbital period of GJ 581b and the period of the variation, we tested the possibility that the variation is due to scattered light from the surface of the planet. Assuming a conservatively large radius of 0.7 $R_{Jup}$ for GJ 581b, ($R_{P}$/$a$)$^{2}$ gives a signal with a semi-amplitude of approximately 0.006$\%$ or 0.07 mmag. This is much smaller than the semi-amplitude of the variation we observe. Thus, the variation is not (solely) due to scattered light. Given the long-term activity trend in the host star (indicating the presence of a magnetic field) observed by \cite*{Gom11,Gom12}, another possible explanation is that the planet induces a spot on the stellar surface, which remains synchronized with the planet's orbital motion. The fact that the 2007 photometry does not clearly show a similar strength signal at the same period may be related to the magnetic activity cycle detected by \cite{Gom12}.

The most convincing case of magnetic star-planet interactions so far is the Tau Bootis system. A periodicity close to the orbital period of the planetary companion was detected in MOST photometry of the system \citep{Wal08}, indicating the possible existence of stellar spots and/or active regions induced by Tau Bootis b. Spectropolarimetric observations suggest the magnetic polarity reversal may be at least partly related to the presence of the planet (\citealt{Don08}; \citealt{Far09}). GJ 581b ($M_{P}\sin i = 0.05 M_{Jup}$; \citealt{For11}) is less massive than Tau Bootis b ($M_{P}\sin i = 3.9 M_{Jup}$; \citealt{But97}). However, the same can be said for the host stars ($M_{\star} = 0.3 M_{\odot}$ and $1.3  M_{\odot}$ for GJ 581 and Tau Bootis, respectively). Further, the b planet is also the largest in the GJ 581 system. Thus, star-planet interactions are a viable possibility but further monitoring of the GJ 581 system is necessary to better understand the nature of the photometric variability observed in the 2009 MOST photometry.

\begin{figure}[!h]
\begin{flushleft}
\includegraphics[scale=0.215]{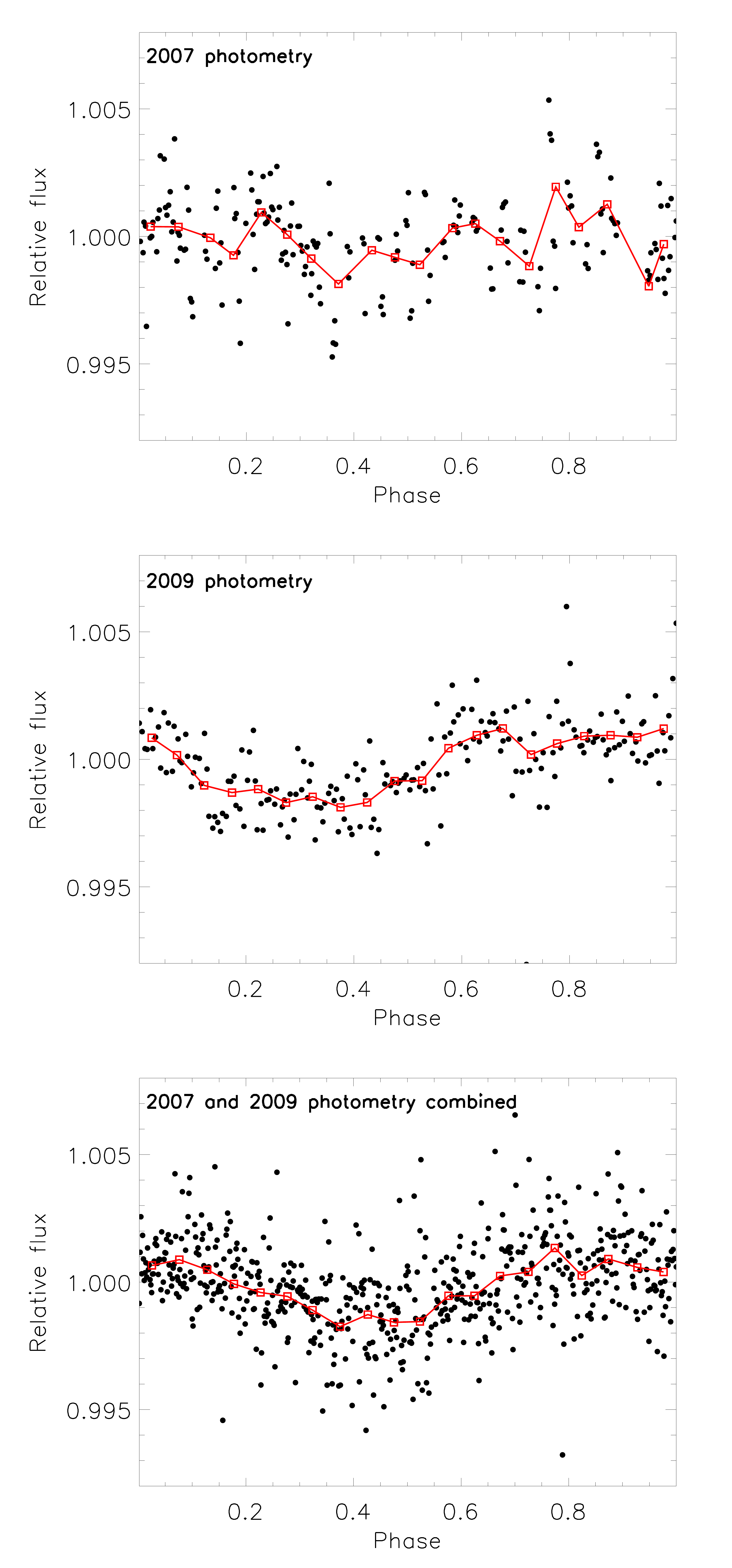}
\caption{The 2007 and 2009 photometry phased individually (top and middle panels) and combined (bottom panel) at the orbital period of GJ 581b. The black solid points represent the photometry binned every 0.005 orbital phase. The red open squares correspond to the data averaged into 0.05 phase bins. }
\end{flushleft}
\end{figure}

%%%%%%%%%%%%%%%%%%%%%%%%

\section{Conclusions}

We observed GJ 581 with the MOST space telescope for a duration of 28 days in 2007 and 11.5 days in 2009.  The 2007 photometry served to determine the level of variability of the host star, while the higher-cadence 2009 photometry was collected with the purpose of searching for transits of GJ 581e. We computed 3$\sigma$ transit windows using both the Keplerian and circular ephemerides reported in \cite{For11}. To quantify the sensitivity of our data to super-Earth transits of various depths, we performed Monte Carlo simulations which involved injecting simulated transits into the light curve and attempting to recover them using a box-shaped model. The simulated transits have planetary radii between 1 and 4 $R_{\oplus}$ and impact parameter values ranging from 0.0 to 0.9.
 
Our transit search returns no significant candidate. This is in agreement with the findings of \cite{For11}. However, our data allow a full search of the 3$\sigma$ circular orbit transit window, and the 3$\sigma$ Keplerian orbit window as well, ensuring that the potential transit of GJ 581e (if sufficiently deep) was not missed by our observations. We place lower limits on the planetary density of GJ 581e, if it transits. If it does not, then its mass likely lies between 1.84 M$_{\oplus}$ and 3.03 $M_{\oplus}$ (assuming orbital co-planarity and stability over more than a few Myr for planets b, c, d and e) as found by \cite{May09}, but we cannot constrain its density.

We conclude that if GJ581e has an envelope dominated by water, helium and/or hydrogen (as appears to be the case for GJ 1214b and Kepler-11f), we rule out transits for most geometric configurations ($b \leq$ 0.6). We cannot rule out transits for a higher density planet. Given the significance of this planetary system, we recommend that RV measurements continue to be collected in order to constrain the eccentricity of the orbits more precisely. Subsequently, further photometric observations should be undertaken to conclusively determine whether GJ 581e, if it has a density higher than 2.5-3 g cm$^{-3}$, transits. A positive result would provide much needed insight into the bulk and atmospheric composition of this low-mass super-Earth.

Since the MOST photometry spans two orbital cycles of GJ 581b, we are able to rule out transits for all plausible compositions of this planet, but only for low impact parameter ($b\lessapprox0.4$). This leaves a significant range of unexplored transiting configurations. We recommend the same course of action as for planet e: once the orbital parameters of the system are more tightly constrained, combining existing GJ 581 photometry with a few hours of new strategically acquired photometry should allow a definitive conclusion to be drawn regarding the presence or absence of GJ 581b transits. 

Transit searches for GJ 581c and GJ 581d cannot be adequately carried out with the photometry presented in this paper, though we are able to place a very loose constraint on the size of GJ 581c, if it transits, in the form of an upper limit of $R_P=2.3 R_{\oplus}$.

By combining the 2007 and 2009 photometry, we find that GJ 581 is stable to within 1\%, indicating that it is likely a quiet M dwarf and supporting evidence of its advanced age ($\gtrsim$ 7 Gyr; \citealt{Sel07}). This increases the probability that life may occur and evolve on a planet in the habitable zone of the system. We attempt to verify the stellar rotation period of 94.2 days determined by \cite{Vog10}, but our data set does not cover a sufficient portion of a rotation cycle of this length to allow any definite conclusions to be drawn. Finally, in the 2009 data set we detect a signal with a period of 5.586 $\pm$ 0.051 days, which is close to the orbital period of GJ 581b. A signal at the same period and of similar strength is not present in the 2007 photometry. We phase the two data sets at the period of the planet, individually and combined, and plot them in Figure 10. Though a signal is not apparent in the 2007 light curve, one is clearly visible in the 2009 and the combined light curves. Additional observations are needed to shed more light on the nature of this possible phase variation.

%%%%%%%%%%%%%%%%%%%%

\section{Acknowledgments}

The authors would like to thank Micha\"{e}l Gillon, Stephen Kane and Bryce Croll for useful conversations. We are grateful to Xavier Bonfils and Thierry Forveille for providing additional information regarding their analysis of the GJ 581 system parameters. We also thank the anonymous referee for valuable comments and suggestions which have helped improve and clarify the manuscript. D.D. is supported by a University of British Columbia Four Year Fellowship. The Natural Sciences and Engineering Research Council of Canada supports the research of DBG, JMM, AFJM and SMR. Additional support for AFJM comes from FQRNT (QuŽbec). RK and WWW were supported by the Austrian Science Fund (P22691-N16) and by the Austrian Research Promotion Agency-ALR.

\end{document}